\documentstyle[prl,aps,epsf,float,twocolumn]{revtex}
\begin{document}
\draft
\twocolumn[\hsize\textwidth\columnwidth\hsize\csname@twocolumnfalse\endcsname
\title{High Magnetic Field ESR in the Haldane Spin Chains NENP and NINO}
\author{M.Sieling$^{(1)}$, U.L\"{o}w$^{(2)}$, B.Wolf$^{(1)}$,
        S.Schmidt$^{(1)}$, S. Zvyagin$^{(1)}$ and B.L\"{u}thi$^{(1)}$}
\address{$^{(1)}$Physikalisches Institut, Universit\"{a}t Frankfurt,
                 D-60054 Frankfurt, Germany\\
         $^{(2)}$Universit\"{a}t Dortmund, D-44221 Dortmund, Germany  }
\date{\today}
\maketitle

\begin{abstract}
We present electron spin resonance experiments in the one-dimensional
antiferromagnetic $S=1$ spin chains NENP and NINO in pulsed magnetic fields up
to 50~T. The measured field dependence of the quantum energy gap for $B||b$ is
analyzed using the exact diagonalization method and the density matrix
renormalization group method (DMRG). A staggered anisotropy term
$(-1)^{i}d(S_i^xS_i^z+S_i^zS_i^x)$ was considered for the first time in
addition to a staggered field term $(-1)^{i}S_i^xB_{\mbox{st}}$. We show that
the spin dynamics in high magnetic fields strongly depends on the orthorhombic
anisotropy~$E$.
\\
\end{abstract}
 ]
\newpage

According to the Haldane conjecture\cite{Haldane} the groundstate properties
and the excitation spectrum of a one--dimensional Heisenberg antiferromagnet
are determined by the spin value $S$ in a fundamental way. For integer spins
the system exhibits a disordered nonmagnetic groundstate which is separated
from the first excited triplet states by an energy gap, whereas half--integer
spin systems have no gap. The existence of the Haldane gap has been
experimentally confirmed by inelastic neutron scattering (INS)
experiments\cite{Renard87} and magnetization measurements\cite{Katsumata}.

Electron spin resonance (ESR) experiments have proven to be an excellent method
to study the magnetic field dependence of the excitation spectrum in Haldane
systems\cite{Date} --\cite{Hagiwara}. In the prototypical one--dimensional spin
chains Ni(C$_{2}$H$_{8}$N$_{2}$)$_{2}$NO$_{2}$(ClO$_{4}$) (NENP) and
Ni(C$_{3}$H$_{10}$N$_{2}$)$_{2}$NO$_{2}$(ClO$_{4}$) (NINO) transitions between
excited states have been observed at the Brillouinzone boundary at $q=\pi / b$
\cite{Date,Brunel}. In addition transitions from the singlet groundstate at
$q=0$ to the excited triplet at $q=\pi / b$ have been
found\cite{Lu}--\cite{Hagiwara}, which are normally forbidden by selection
rules. These ESR experiments could be explained using the concept of a
transverse staggered magnetic field\cite{Shiba,Mitra} which mixes the
eigenstates at $q=0$ and $q=\pi / b$. Such a staggered field could be induced
by an external magnetic field due to the alternating tilting of the local
anisotropy axes of the Ni$^{2+}$ ions and was first observed in NMR
experiments\cite{Chiba}. ESR experiments are so far the only method for a
direct observation of the energy gap in Haldane systems well above the critical
field $B_c$ (8.5~T in NENP and $\approx 8$~T in NINO).

NENP and NINO both crystallize in the orthorhombic system and show similar
lattice constants\cite{Renard88}~--~\cite{Yosida}. NENP can be described in the
$Pnma$ space group\cite{Meyer} whereas NINO belongs to the $Pbn2_1$ space
group\cite{Renard88,Yosida}. The Ni$^{2+}$ ions are covalently linked via
NO$_2$--complexes which are responsible for the antiferromagnetic superexchange
interaction. Both compounds are good realizations of one--dimensional chains
since the interchain exchange $J'$ is about a factor of $10^4$ smaller than the
intrachain exchange $J$\cite{Renard88} due to the separation  by the ClO$_4$
ions. The local surrounding of the Ni$^{2+}$ ions is a distorted octahedron
with a basal plane built out of four nitrogen atoms from the diamine groups.
The axial positions are occupied by an oxygen and a nitrogen atom from the
nitrite bridges. This local surrounding gives rise to a planar anisotropy $D$
and to an orthorhombic anisotropy $E$ as well as to an anisotropic $g$~tensor.

A significant characteristic in NENP and NINO is the alternating tilting of the
local surrounding  of the Ni$^{2+}$ ions along the chain. As we will show in
this paper, it is important to know the direction of this tilting with respect
to the easy direction of the orthorhombic anisotropy $E$. For NENP it is clear
from the structure\cite{Meyer} and NMR experiments\cite{Chiba} that the tilting
of the local environment is along the $c$ direction \cite{footnote} while the
easy direction of the orthorhombic anisotropy is along $a$ (see e.g.
\cite{Katsumata}). In NINO the easy direction in the basal octahedron plane is
along $c$\cite{Takeuchi} while the structure \cite{Yosida} gives rise to the
assumption that the local environment is tilted towards the $a$-axis. However
it is not clear how to deduce a definite angle of tilting of the $g$ or the $D$
tensors out of the structure.

The ESR experiments were performed in the high magnetic field laboratory at the
University of Frankfurt. Different single crystals of NENP and NINO have been
mounted in a transmission type sample holder in Voigt geometry (propagation
vector $\vec{k}$ perpendicular to the external magnetic field $\vec{B}$) with
the chain direction $b$ parallel to $\vec{B}$. We used Gunn and IMPATT diodes
together with frequency multipliers to produce mm-wave radiation in the
frequency range 55--200~GHz. The radiation was unpolarized when passing the
sample and was detected with a fast InSb hot electron bolometer mounted in a
separated cryostat. High magnetic fields were produced using an Al$_2$O$_3$
strengthened copper coil with internal reinforcement, manufactured in the
National High Magnetic Field Laboratory in Tallahassee, Florida. Fields up to
50~T with 8~ms rise time could be produced by discharging capacitors with a
stored energy of 0.4~MJ.
\begin{figure}
\centerline{\epsfxsize=3.3in\epsffile{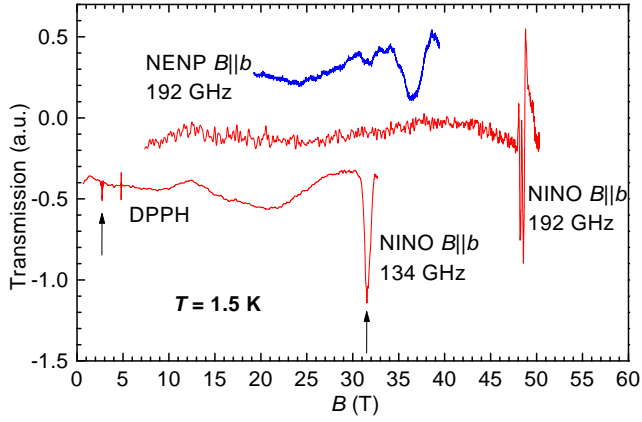}} \caption{Transmission
spectra in NENP and NINO at $T=1.5$~K. The arrows mark excitations from the
groundstate below and above the critical field in the 134~GHz spectrum.}
\label{fig1}
\end{figure}
While the coil is operating at 77~K, the sample was cooled down to 1.5~K in a
pumped $^4$He cryostat. The magnetic field was measured with a pick up coil and
calibrated using a DPPH reference sample. Details of the high field setup can
be found in \cite{Wolf}

Figure~\ref{fig1} shows some typical mm--wave transmission spectra at $T$=1.5~K
for NENP and NINO in high magnetic fields. A drastic increase of the resonance
intensity can be observed with increasing field above the critical field $B_c
\approx 8~\mbox{T}$ (see corresponding arrows in fig.~\ref{fig1}). This
increase of intensity was explained before in the framework of the staggered
field concept\cite{Shiba,Mitra}. Also the linewidth increases remarkably in
both substances: while the typical linewidth for $B<B_c$ in NENP and NINO is
about 0.1~T and 0.05~T respectively, it increases for $B>B_c$ about a factor of
20 in NENP and a factor of 10 in NINO. No decrease of the integrated intensity
could be observed with decreasing temperatures, indicating that the observed
transitions are excitations from the groundstate\cite{Sieling}. The resonance
frequencies therefore give a direct measure of the magnetic field dependence of
the energy gap\cite{Shiba,Mitra}.

\begin{figure}
\centerline{\epsfxsize=3.3in\epsffile{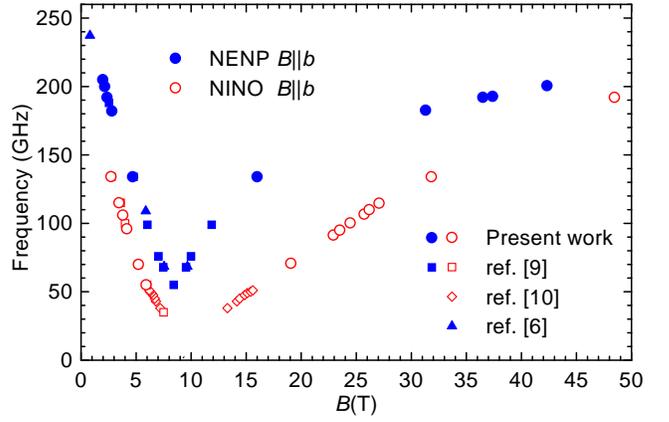}} \caption{Frequency field
dependence of the groundstate excitations in NENP and NINO (full and open
symbols respectively).} \label{fig2}
\end{figure}

The frequency field dependence for all observed resonances is shown in
fig.~\ref{fig2} together with low field data obtained by other groups. For
$B<B_c$ the gap in both systems decreases with a slope corresponding to the
g~factor ($g_b=2.15$ in NENP, $g_b=2.17$ in NINO). The gap does not close but
remains finite at the critical field $B_c$ and increases again for $B>B_c$. For
$B>B_c$ the curvature of the $\Delta(B)$ curve is much stronger in NENP than in
NINO. The smaller slope in the high field region for NENP seems to be one
explanation for the larger linewidth in this compound.

\begin{figure}
\centerline{\epsfxsize=3.3in\epsffile{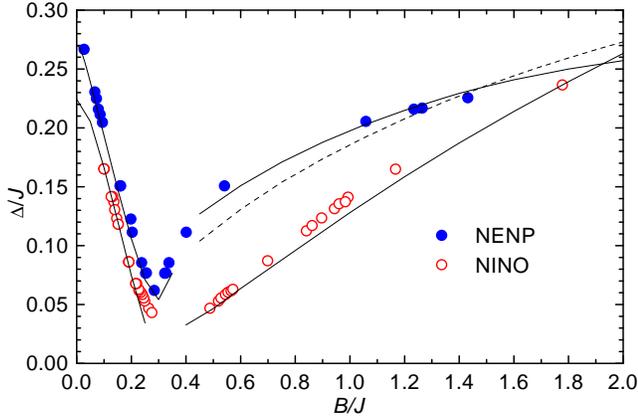}} \caption{Magnetic field
dependence of the gap in NENP and NINO in units of $J$. Symbols are
experimental data, lines are extrapolated numerical results based on the
Lanczos method (for $B/J < 0.4$) and the DMRG method (for $B/J > 0.4$). The
staggered field and anisotropy parameters are $c$=-0.026, $d/J$=0.06 and
$c$=0.0105, $d/J$=0.06 for NENP and NINO respectively (solid lines). The dashed
line is the best fit for NENP with $d/J=0$ using $c=0.017$.} \label{fig3}
\end{figure}

The difference in the high field spin dynamics between NENP and NINO was
thought to be mainly caused by a larger staggered inclination of the Ni$^{2+}$
surrounding in NENP\cite{Sieling,Hagiwara} (although these assumptions are
based on calculations with an isotropic Hamiltonian and unrealistically small
exchange constants $J$). Thus the ratio of
the gap energies and the ratio of the slope of the $\Delta(B)$ curves are
expected to be approximately constant between $B_c$ and $B=4J$\cite{Shiba}. In
contrast the actual field dependence of the gap energies shows a different
behaviour with a crossing point of both $\Delta(B)$ curves at $B=1.8J$ (fig
\ref{fig3}).

To analyze our data for the $B || b$ configuration more realistically we use
the following Hamiltonian
\begin{eqnarray}\label{Hamilton}
  \lefteqn{H=\sum_i \{ J \vec{S}_i \vec{S}_{i+1} + D (S_i^z)^2
             + E[(S_i^x)^2 - (S_i^y)^2]} \nonumber \\
& & {} - g_z\mu_{\text{B}}S_i^z B + (-1)^i[d(S_i^x S_i^z + S_i^z S_i^x) + c
g_z\mu_{\text{B}} S_i^xB] \} \mbox{ .}
\end{eqnarray}
The indices $x$, $y$ and $z$ correspond to the crystallographic axes $c$, $a$
and $b$ respectively. $J$ is the antiferromagnetic intrachain exchange, $D$ and
$E$ are the planar and the orthorhombic anisotropies. The fourth and the
following terms contain the Zeeman coupling and the {\it staggered\/} terms due
to an alternating tilting of the anisotropy- and the $g$-tensor within the
$x$--$z$ plane. $c$ is the staggered field constant $c=B_{\text{st}}/B$ which
is given by the tilting angle $\theta_g$ and the difference of $g_x - g_z$:
\begin{equation}\label{c}
  c = \frac{g_x - g_z}{g_z}\frac{1}{2}\tan2\theta_g\mbox{ .}
\end{equation}
$d$ is the staggered anisotropy constant which is given by $\theta_D$ and the
difference of $D-E$:
\begin{equation}\label{d}
  d = (D-E)\frac{1}{2}\tan2\theta_D\mbox{ .}
\end{equation}
The parameters $D$, $E$, $g_x$ and $g_z$ are the observable values in the
crystallographic system. They can be derived by averaging over the tilted
anisotropy- and $g$-tensors of neighboring Ni$^{2+}$ sites. For a more detailed
derivation of (\ref{c}) and (\ref{d}) we refer to ref.~\cite{Sagi}.

We like to point out that although the Hamiltonian (\ref{Hamilton}) contains
numerous parameters, only the angles~$\theta_g$ and $\theta_D$ can be
considered as "free" parameters within the fitting process for our high field
data. $J$, $D$ and $E$ are fixed by
the zero field size of the gaps $\Delta_1$, $\Delta_2$ and $\Delta_3$ between
the groundstate and the excited triplet states at $q=\pi/b$. $J$ is given by
the well known ratio $0.41J=E_{\text{G}}=(\Delta_1 + \Delta_2 + \Delta_3)/3$
\cite{Nightingale} while $D$ and $E$ can be deduced from the splitting of the
triplet $\Delta_3 - (\Delta_1 + \Delta_2)/2$\cite{Sakai} and $\Delta_2 -
\Delta_1$ respectively. The gap values are taken from ESR experiments
\cite{Sieling} which are in good agreement with INS results \cite{Renard87}.
The $g$-factors for the different directions are taken from zero field
susceptibility data (\cite{Meyer} for NENP and \cite{Takeuchi} for NINO).

Accordingly we obtain the following parameters $J$=43~K, $D/J$=0.17,
$E/J$=0.01, $g_x$=2.21, and $g_z$=2.15 for NENP and $J$=39~K, $D/J$=0.2,
$E/J$=-0.027, $g_y$=2.23, and $g_z$=2.17 for NINO. As noted before, in NINO the
easy direction of the orthorhombic anisotropy is along the $c$- ($x$-) axis
leading to a negative sign of $E$. Since the alternating tilting in NINO is
supposed to be in the $y$-$z$ plane the $S_i^x$ operators and $g_x$ in the
staggered term of (\ref{Hamilton}) have to be replaced by $S_i^y$ operators and
$g_y$ respectively.

The magnetic field dependence of the gap was calculated both with exact
diagonalization for $N$=16 spins using the Lanczos algorithm and with the
density matrix renormalization group (DMRG) method. The Lanczos results for
short chains can be extrapolated in a reliable way for $B<B_c$. For  $B>B_c$
however the gap shows large finite size effects which manifest themselves in
oscillations. This is in particular the case for small values of the staggered
fields, when both the ground state energy and the first excited state energy
follow almost polygonal lines as a function of B. Therefore for $B>B_c$ we use
the DMRG method which allows to treat large systems with high accuracy by
keeping only the most important eigenstates of a blocked density matrix
constructed according
\begin{figure}
\centerline{\epsfxsize=3.3in\epsffile{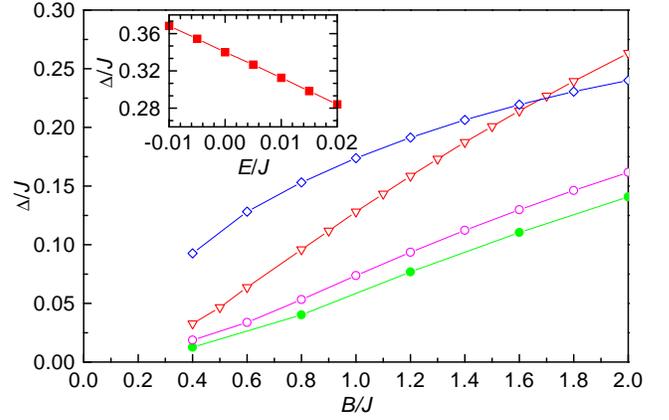}} \caption{Magnetic field
dependence of the high field energy gap for different parameters. Full circles
($\bullet$) are calculation for $c=0.0105$, $d/J=0$, $D/J=0.2$ and $E/J=0.027$.
Open symbols are calculations which differ from ($\bullet$) in only one
parameter (($\circ$) $c=0.012$, ($\bigtriangledown$) $d/J=0.06$, ($\diamond$)
$E/J=0$). The inset demonstrates the linear dependency of the gap on the
orthorhombic anisotropy $E$ at $B/J=2$ for $c=0.0105$, $d/J=0.06$ and $D=0.2$.
All results are extrapolations of DMRG calculations.
 }\label{fig4}
\end{figure}
to ref.~\cite{White}. We found best convergence using DMRG for $N=40\dots120$
spins with open boundaries but also employed periodic boundary conditions in
order to remove the $S=1$ artefact state occurring for open boundaries in the
low field range. In the high field region the method converges very well with
as little as 30 states. In the region near the critical field more states and
larger systems are needed.

Figure \ref{fig3} shows the extrapolated results of our calculations together
with the experimental data in units of $J$. For NINO we find $c=0.0105$ and
$d/J=0.06$ assuming a tilting angle of $\theta_g=\theta_D=17.5^{\circ}$. For
NENP we obtained our best fit for $c=-0.026$ and $d/J=0.06$ based on the
tilting angles $\theta_g = -31^{\circ}$ and $\theta_D =
18.5^{\circ}$\cite{angle} which differ in size and direction.

This somewhat strange result implies that the the $g$- and the $D$- tensors in
NENP depend on different neighboring atoms of the Ni$^{2+}$ ions, e.g. one
tensor is mainly determined by the position of the nitrogen atoms of the
ethylenediamine groups whereas the other tensor is principally given by the
position of the nitrite bridges. A more detailed analysis is needed at this
point to clarify the dependence of the $g$- and $D$-tensors on the Ni$^{2+}$
environment. Nevertheless neglecting the staggered $d$ term leads to a rather
poor fit for the new high field data of NENP (dashed line in fig.~\ref{fig3})
showing that this $d$ term is essential to describe the gap as a function of
the field. Besides one has to keep in mind that the interchain exchange $J'$
was not taken into account in our calculations which possibly could affect the
high field spin dynamics.

It is of special interest to study the effect of the different terms in
(\ref{Hamilton}) on the energy gap when varying several parameters. In a first
step we checked that finite values of $c$ and $d$ have practically no effect
for $B<B_c$, thus confirming the values taken for $J$, $D$, and $E$. Next we
studied the effect of $c$ and $d$ in the high field range. For $B>B_c$ the
system is expected to be gapless without any staggered term in
(\ref{Hamilton}). The staggered terms lift the degeneracy between the magnetic
groundstate and the first exited state in the high field phase yielding a gap
which increases with $B$. Fig.~\ref{fig4} shows the gap for different $c$ and
$d$ values as a function of field. The effect of the $d$ term on the gap is
quite small at the critical field $B_c \approx 0.4J$ but increases with $B$
comparable with the increase of the magnetization $\sum<S_z>$ reflecting the
effect of the $S^z_i$ operators in (\ref{Hamilton}).

As is well known the anisotropies $D$ and $E$ reduce the gap for $B<B_c$
\cite{Nightingale}, \cite{Sakai}. While the planar anisotropy $D$ only has a
very weak effect for $B>B_c$ it turned out that the orthorhombic anisotropy $E$
considerably changes the size of the high field gap induced by the staggered
$c$ and $d$ terms (fig.~\ref{fig4} with inset). The gap is linearly reduced for
a positive $E$ in (\ref{Hamilton}) whereas it is linearly enlarged for $E<0$
(see inset of fig.~\ref{fig4}). The staggered $c$ and $d$ terms induce a
staggered magnetization $\sum(-1)^i <S_i^x>$ along the $x$ direction in NENP
\cite{Shiba} (along the $y$ direction in NINO), which is the hard axis of the
orthorhombic anisotropy respectively. Thus a positive $E$ increases the energy
of the groundstate and the first excited states. This increase is somewhat
stronger for the groundstate which shows a higher staggered magnetization than
the first excited states yielding a reduction of the gap.

The orthorhombic anisotropy term is essential to describe the differences in
the high field spin dynamics of NENP and NINO. As shown in fig.~\ref{fig4} the
field dependence of the gap has a clear curvature for $E=0$ while it becomes
more linear with increasing $E$ corresponding to the experimental results in
both substances.

In the present work we have shown unprecedented data for the quantum energy gap
in the Haldane systems NENP and NINO measured by ESR experiments in pulsed
magnetic fields up to 50~T. We presented numerical calculations based on a
direct diagonalization and the DMRG method considering the planar and
orthorhombic anisotropies $D$ and $E$ as well as the staggered anisotropy and
the transverse staggered field. The high field data could be fitted using
realistic values for the exchange interaction and the anisotropies. In addition
to a staggered field term the staggered anisotropy term is essential to
describe the curvature of the gap as a function of field. Finally the
orthorhombic anisotropy $E$ turned out to have an unexpected strong effect on
the energy gap for $B>B_c$.

Future experiments with high magnetic fields perpendicular to the chain axis
could provide us with more information especially for the case where no
staggered field is induced. In addition experimental and theoretical
investigations on the field dependence of higher excited states should extend
the knowledge about the high field phase in these Haldane systems.

We like to thank T.~Yosida and M.~Date who kindly gave us the samples of NENP
and NINO. This work was supported by the BMBF with the project 13N6581A/3 and
by the DFG through the SFB~252. S.Z.~appreciates the support of the Alexander
von Humboldt Foundation.


\begin{references}
\bibitem{footnote} In fact in NENP neighboring chains are rotated about the
                   chain axis $b$ by approximately $\pm30^\circ$ leading to
                   different local and crystallographic symmetry axes. Here we
                   just discuss the $B||b$ configuration and therefore will
                   neglect this rotation.
\bibitem{angle} Note that the angle $\theta$ contains a considerable
error due to the experimental accuracy of the g-factor determination in eq.~2.
\end{references}
\end{document}